\documentclass[aps,pra,reprint,showpacs,floatfix]{revtex4-1}

\usepackage{amsmath}
\usepackage{slashed}
\usepackage{txfonts}
\usepackage{microtype}
\usepackage{graphicx}
\usepackage[
breaklinks=true
]{hyperref}
\usepackage{color}
\usepackage{ulem}
\def\sout{\bgroup\markoverwith
{\textcolor{red}{\rule[0.5ex]{2pt}{0.5pt}}}\ULon}
\def\be{\begin{equation}}
\def\ee{\end{equation}}
\def\bes{\begin{equation*}}
\def\ees{\end{equation*}}
\def\bea{\begin{eqnarray}}
\def\eea{\end{eqnarray}}
\def\beas{\begin{eqnarray*}}
\def\eeas{\end{eqnarray*}}
\def\bal#1\eal{\begin{align}#1\end{align}}
\def\bals#1\eals{\begin{align*}#1\end{align*}}

\renewcommand{\vec}[1]{\mathbf{#1}} 
\newcommand{\veg}[1]{\boldsymbol{#1}} 
\newcommand{\del}{\partial}

\graphicspath{{figures/}}

\renewcommand*{\vec}[1]{\boldsymbol{#1}}

\begin{document}
\title{On the quantization of the nonintegrable phase in electrodynamics}

\author{Enderalp \surname{ Yakaboylu}} \thanks{\mbox{Corresponding
author:enderalp.yakaboylu@mpi-hd.mpg.de}}
\affiliation{Max-Planck-Institut f\"ur Kernphysik, Saupfercheckweg 1, D-69117 Heidelberg, Germany}
\author{Karen Z. \surname{Hatsagortsyan}} 
\affiliation{Max-Planck-Institut f\"ur Kernphysik, Saupfercheckweg 1, D-69117 Heidelberg, Germany}
\date{\today}

\begin{abstract}
Using the fact that the nonintegrable phase factor can reformulate the gauge
theory in terms of path dependent vector potentials, the quantization condition for the
nonintegrable phase is investigated. It is shown that the path-dependent formalism can provide compact description
of the flux quantization and the charge quantization at the existence of a magnetic monopole.
Moreover, the path-dependent formalism gives suggestions for searching of the quantized flux in
different configurations and for other possible reasons of the charge quantization. As an example,
the developed formalism is employed for a (1+1) dimensional world, showing the relationship
between
the fundamental unit of the charge and the fine structure constant for this world.
\end{abstract}


\maketitle

\section{Introduction}

The homogeneous Maxwell equations, Gauss' law for magnetism and Faraday's law of
induction, identifies the electromagnetic field strength tensor $F_{\mu
\nu}$, whose components are the electric and magnetic field, in terms of gauge
dependent vector potentials. Furthermore, any other vector potential, 
related by a so-called gauge transformation, describes the same electric and
magnetic field. Although, all the physical observables are independent from the vector potentials,
hence they are gauge invariant, it is essential to introduce them for the Hamiltonian formulation of
the dynamics. Moreover, as elegantly described by Aharanov and Bohm, the
vector potential has a significant role in quantum mechanics \cite{Aharonov_Bohm_1959}.
The vector potential does not only provide a compact
mathematical formulation of the associated field strength tensor but also it leads to
new predictions such as Aharanov-Bohm effect \cite{Ehrenberg_Siday_1949, Aharonov_Bohm_1959,
Aharonov_Bohm_1961}, flux quantization \cite{Onsager_1953, London_1950, Deaver_Fairbank_1961,
Doll_Nabauer_1961} and the Dirac's charge quantization condition in the existence of a magnetic
monopole \cite{Dirac_1931, Dirac_1948}. Later, in their celebrated paper \cite{Wu_Yang_1975}, Wu and Yang 
gave a complete description of electromagnetism based on the concept of
the nonintegrable phase factor
\be
\label{nonintegrable_phase_factor}
\exp\left( - \dfrac{i e}{\hbar c} \int_{\mathcal{P}}^x A_\nu d y^\nu \right).
\ee
The integration path starts at a point $\mathcal{P}$ where the fields are zero and runs up to the
point of interest $x$.
Wu and Yang pointed out that the field strength tensor underdescribes the complete electromagnetic phenomena, in other words, the different physical realization of electromagnetic phenomena may have the same field strength tensor $F_{\mu \nu}$.
Historically, such kind of line integrals of the potentials have previously been suggested in
\cite{Bergmann_1956, Dirac_1955, DeWitt_1962, Mandelstam_1962}. Moreover, it was shown by DeWitt \cite{DeWitt_1962} and
Mandelstam \cite{Mandelstam_1962} that the nonintegrable phase factor Eq.~(\ref{nonintegrable_phase_factor}) can eliminate
the gauge freedom from the formalism. However, the expense is that the vector potentials, which depend on the field
strength tensor, become path dependent. Every gauge functions in the conventional gauge theory have a
counterpart path in this equivalent formulation \cite{Belinfante_1962, Rohrlich_Strocchi_1965}.

In the present manuscript, we discuss the quantization condition of the nonintegrable phase in
the light of the  path-dependent formalism of the gauge theory (shortly, we will call it the
path-dependent formalism),
and explore topological electromagnetic effects in quantum
mechanics. It is demonstrated that the path-dependent formalism provides a clear description of the
electromagnetic flux quantization which could point out possible reasons for the charge quantization. 
We apply the developed formalism  for a (1+1) dimensional world and find a relationship between the
fundamental unit of the charge and the fine structure constant for this world.

After briefly summarizing the gauge theory in Sec.~\ref{gauge_theory}, we
show explicit derivation of the path-dependent formalism in Sec.~\ref{PDF} and the correspondence
between the conventional gauge theory and the path-dependent formalism
in a pedestrian
level.
The condition on the quantization of the nonintegrable phase is discussed in
Sec.~\ref{quantization_of_phase}, where the flux and the Dirac's charge quantization conditions are
given. In Sec.~\ref{charge_quantization_section}, the quantization of the charge and the estimation of its fundamental
units are illustrated in a (1+1) dimensional world. The conclusion and further remarks are given in
Sec.~\ref{conclusion}. 

The CGS units and the metric convention $g = (+,-,-,-)$ are used throughout the paper.

\section{Conventional Gauge Theory} \label{gauge_theory}

The abelian gauge theory can be summarized in the light of
\cite{Mills_1989}. In classical electrodynamics the Maxwell equations in (3+1) spacetime
dimensions read
\bea
\label{non_hom_me} \del_\mu F^{\mu \nu} &=& \dfrac{4 \pi }{c} J^\nu\, , \\
\label{hom_me}  \epsilon^{\alpha \beta \mu \nu} \del_\mu F_{\alpha \beta} &=& 0\, ,
\eea
where the components of the electromagnetic field strength tensor $F^{\mu \nu}$
are the electric field $F^{i 0} = E^i $ and the magnetic field $F^{jk} = \epsilon^{ijk} B_k
$, and the four-vector current is defined $j^\mu = (c \rho, \vec{J})$. 
The homogeneous Maxwell equation (\ref{hom_me})
allow to express 
the  electric and magnetic fields in terms of a four-vector potential $A^\mu =
(\phi, \vec{A})$ as
\be
F_{\mu \nu} =  \del_\mu A_\nu - \del_\nu A_\mu\, . 
\ee
Furthermore, any other four-vector potential, related by a so-called gauge
transformation, describes the same electric and magnetic field. In other words,
the transformation
\be
\label{gauge_trans}
A^\mu \rightarrow A^\mu + \del^\mu \chi
\ee
leaves the electromagnetic field strength tensor invariant and, consequently, all
the physically measurable quantities related to the electrodynamics such as the
Maxwell equations, the Lorentz force law become gauge invariant.

More fundamentally, the gauge invariance can appear as a consequence of the conservation of the electric
charge under local symmetry transformation via Noether's theorem. In quantum theory, the
conservation of the electric charge follows from the local phase invariance of the wave
function. Further, the local phase invariance of the wave function imposes an
interaction between the associated conserved quantity and the gauge field such
that the Schr\"{o}dinger equation which governs the time evolution of the wave
function becomes invariant under the gauge-transformation. For instance, the
Dirac equation for a relativistic spin-$1/2$ particle
\be
\left[ i \hbar \gamma^\mu \left( \del_\mu - \dfrac{i e}{\hbar c} A_\mu \right) -
mc \right] \psi(x) = 0
\ee
is invariant under the transformations (\ref{gauge_trans}) as long as the
wave function transforms as
\be
\psi(x) \rightarrow \exp\left(\dfrac{i e \chi}{c \hbar}\right)
\psi(x)\, .
\ee

In general, the gauge theory can be patterned as follows: First, for every
conservation law there is an associated symmetry via Noether's theorem; second,
the local ones among them lead to the existence of gauge fields; and third, the
gauge field theory imposes interactions between the gauge field and the
conserved quantity. Such an generalization of the local gauge invariance leads to, for instance, the
existence of the non-abelian gauge field \cite{Yang_Mills_1954}.

\section{The path-dependent formalism} \label{PDF}

As it was discussed in \cite{Wu_Yang_1975}, the fundamental concept which
describes
complete electromagnetism
is the nonintegrable phase factor (\ref{nonintegrable_phase_factor}).
The nonintegrable phase factor can eliminate the vector
potential from the formalism \cite{DeWitt_1962}, \cite{Mandelstam_1962}. In fact, let us define the
gauge function $\chi$ via the path integral
\be
\chi = - \int_{\mathcal{P}}^x A_\nu d y^\nu \, ,
\ee
then the associated Schr\"{o}dinger equation becomes invariant under the
following gauge transformation
\be
\label{gauge_invariant_potential}
A_\mu (x) \rightarrow \mathcal{A}_\mu(x) \equiv A_\mu(x) - \dfrac{\del}{\del x^\mu}
\int_{\mathcal{P}}^x
A_\nu d y^\nu \, .
\ee
The latter yields to the gauge invariant vector potential $\mathcal{A}_\mu(x)$. Explicitly, if one parametrizes the
path $y = y(s,x)$ as
\be
\label{boundry-conditions} y(1, x) = x\, , \quad
y(0, x) = x_0 \, ,
\ee
where the electromagnetic field vanishes at $x_0$, at which $A_\mu$ may, without
loss of generality, be set equal to zero. Then, Eq.~(\ref{gauge_invariant_potential}) becomes
\begin{widetext}
 \bea
\nonumber \mathcal{A}_\mu(x) && = A_\mu(x)  - \del_\mu \int_{0}^{1} A_\nu
(y)
\dfrac{\del y^\nu}{\del  s} ds  = A_\mu(x) - \int_{0}^{1}  \left(
\dfrac{A_\nu(y)}{\del y^\lambda} \dfrac{\del y^\lambda}{\del x^\mu} \dfrac{\del
y^\nu}{\del  s} + A_\nu(y) \dfrac{\del }{\del  s} \dfrac{\del
y^\nu}{\del x^\mu} \right)d  s \, , \\
\label{new_vector_potential_2} && = A_\mu(x) - \int_{0}^{1}  \left(
\dfrac{A_\lambda (y)}{\del y^\nu} \dfrac{\del y^\nu}{\del  s} \dfrac{\del
y^\lambda}{\del x^\mu}  + A_\nu(y) \dfrac{\del }{\del  s} \dfrac{\del
y^\nu}{\del x^\mu} - F_{\nu \lambda}(y)  \dfrac{\del y^\lambda}{\del x^\mu}
\dfrac{\del y^\nu}{\del  s} \right)d  s\, ,
\eea
\end{widetext}
where in last line we have used $A_{\nu, \lambda}(y) = A_{\lambda,\nu}(y) -
F_{\nu \lambda}(y)$. Further, the first two integrand terms in Eq.
(\ref{new_vector_potential_2}) can be
written as $ \displaystyle \dfrac{\del}{\del  s}\left(A_\lambda(y) \dfrac{\del
y^\lambda}{\del x^\mu } \right) $ and using the boundary conditions
(\ref{boundry-conditions}),
\begin{subequations}
  \label{path_dependent_potential}
\be
\label{path_dependent_potential_a}
\mathcal{A}_\mu(x) = \int_{0}^{1}  F_{\nu \lambda}(y) \dfrac{\del
y^\nu}{\del s}  \dfrac{\del y^\lambda}{\del x^\mu}  d  s
\ee
is obtained. Furthermore, since the filed strength tensor $F_{\mu \nu}$ is antisymmetric,
Eq.~(\ref{path_dependent_potential_a}) can be written as
\be
\label{path_dependent_potential_b}
\mathcal{A}_\mu(x) = \frac{1}{2} \int_{0}^{1}  F_{\nu \lambda}(y) \left( \dfrac{\del
y^\nu}{\del s}  \dfrac{\del y^\lambda}{\del x^\mu} - \dfrac{\del
y^\lambda}{\del s}  \dfrac{\del y^\nu}{\del x^\mu} \right) d  s \,.
\ee
\end{subequations}
The expression (\ref{path_dependent_potential}) is gauge
independent because it is written solely in terms of the gauge invariant field strength tensor
$F_{\mu \nu}$. However, the expense is that the vector potential is path dependent and
every gauge functions in the conventional gauge theory have a counterpart in the path-dependent
formalism \cite{Belinfante_1962, Rohrlich_Strocchi_1965}. As a consequence, we will
label both the vector potential $\mathcal{A}_\mu$ and the wave function $\Psi$
with the path index $\mathcal{P}$ which refers to a certain path as
\be
\label{path_dep_dirac_eq}
\left[ i \hbar \gamma^\mu \left( \del_\mu - \dfrac{i e}{\hbar c} \mathcal{A}_\mu
(\mathcal{P},x) \right) -
mc \right] \Psi(\mathcal{P},x) = 0\, .
\ee
Moreover, the Dirac equation (\ref{path_dep_dirac_eq}) is invariant under the following path
transformation
\be
\label{path_transformation}
\mathcal{A}_\mu (\mathcal{P'},x) = \mathcal{A}_\mu (\mathcal{P},x) + \del_\mu
\oint_{\del \Sigma}^x A_\nu dy^\nu\, ,
\ee
as long as the wave function satisfies
\be
\Psi[\mathcal{P'},x] = \exp\left(\dfrac{i e}{\hbar c} \oint_{\del \Sigma}^x A_\mu
dy^\mu \right)
\Psi[\mathcal{P},x]\, ,
\ee
with the closed loop ${\del \Sigma} = \mathcal{P}-\mathcal{P'}$. Furthermore,
using the four-dimensional Stokes' law, the loop integral can be converted to surface integral
\be
\label{wilson_loop}
\oint_{\del \Sigma}^x A^\mu d y_\mu = \dfrac{1}{2} \int_\Sigma^x F^{\mu \nu} d
\sigma_{\mu \nu} = \Phi_{EM}(x) \, ,
\ee 
with the electromagnetic flux $\Phi_{EM}$ \cite{Cabibbo_Ferrari_1962}. In addition to that, using
the definition of the path dependent vector potential~(\ref{gauge_invariant_potential}), the
electromagnetic flux for a nonconfined field can also be identified as
\be
\label{flux_via_open_int}
\int_\mathcal{P}^x \mathcal{A}_\mu (\mathcal{P'}) d y^\mu = \Phi_{EM}(x) \, ,
\ee
which implies that for any path $\mathcal{P}$
\be
\label{vanishing_path_integral}
\int_\mathcal{P}^x \mathcal{A}_\mu (\mathcal{P}) d y^\mu = 0 \,
\ee
always holds \footnote{For confined fields, the electromagnetic flux is always given by
Eq.~(\ref{wilson_loop}), and further $\int_\mathcal{P}^x \mathcal{A}_\mu (\mathcal{P}) d y^\mu$ may
not be equal to zero, see Sec.~\ref{quantization_of_phase}}. In conclusion, true electromagnetism
can be described by path invariance of the so-called nonintegrable phase factor which is known as
Wilson loop when the gauge group is nonabelian \cite{Wilson_1974}.

In order to illustrate the equivalence between the conventional gauge theory and the 
path-dependent formalism, let us specify some certain paths which have a well-known
counterpart in the conventional gauge theory. For the sake of simplicity, assume
that there is only a constant and uniform electric field $\vec{E}_0$ and
further,  without loss of generality, let the initial point be $x_0^\mu = (0,
\vec{0})$. If one chooses the path $\mathcal{P} = \mathcal{P}_1 + \mathcal{P}_2$
with the each segments
\bea
\mathcal{P}_1: && \quad y^\mu (s, x) = (0, s \, \vec{x})\, , \quad 0 \le s \le 1 \, ,
\\
\mathcal{P}_2: && \quad y^\mu (s, x) = (s \, ct, \vec{x})\, , \quad 0 \le s \le 1 \, , 
\eea 
then path dependent vector potential becomes
\bea
\nonumber \mathcal{A}^\mu(x) &=& F_{0i} \int^1_0 \dfrac{\del y^0}{\del s} \dfrac{\del
y^i}{\del x_\mu} ds, \\
 &=& \left( 0, - ct \, \vec{E}_0 \right). 
\eea
This gauge is called velocity gauge. On the other, if we
choose the segments of the path as
\bea
\mathcal{P'}_1: && \quad y^\mu (s, x) = (s \, ct, \vec{0})\, , \quad 0 \le s \le 1 \, ,
\\
\mathcal{P'}_2: && \quad y^\mu (s, x) = (ct, s \, \vec{x})\, , \quad 0 \le s \le
1, 
\eea
then the vector potential yields
\bea
\nonumber \mathcal{A}^\mu(x) &=& F_{i0} \int^1_0 \dfrac{\del y^i}{\del s} \dfrac{\del
y^0}{\del x_\mu} ds \, , \\
 &=& \left( - \vec{x} \cdot \vec{E}_0, \vec{0} \right)
\eea
which is know as length gauge. Furthermore, if we trace a
straight line as
\be
\mathcal{P''}: \quad y^\mu (s, x) = (s\, ct, s \, \vec{x})\, , \quad 0 \le s \le 1 \, ,
\ee
then, the vector potential is given by
\bea
\nonumber \mathcal{A}_\mu(x) &=& F_{i0} \int^1_0 \left( \dfrac{\del y^i}{\del s}
\dfrac{\del y^0}{\del x^\mu} - \dfrac{\del y^0}{\del s} \dfrac{\del y^i}{\del
x^\mu}  \right) ds\, , \\
\mathcal{A}^\mu (x) &=& \left(- \dfrac{1}{2} \vec{x} \cdot \vec{E}_0,  -
\dfrac{1}{2} ct \, \vec{E}_0 \right)\, ,
\eea
which is know as Fock-Schwinger gauge $x_\mu \mathcal{A}^\mu = 0 $. Following
the path
transformation (\ref{path_transformation}), relation between different gauges is
given by the electromagnetic flux. For instance, gauge transformation between
velocity and length gauge is given by
\be
\Phi_{EM}(x) =  - ct \, \vec{x} \cdot \vec{E}_0
\ee
with $\del \Sigma = \mathcal{P} - \mathcal{P'}$. Similarly, transformation
between the length gauge and the Fock-Schwinger gauge can be accomplished by
\be
\Phi_{EM}(x) =  - \dfrac{1}{2} ct \, \vec{x} \cdot \vec{E}_0
\ee
with $\del \Sigma = \mathcal{P'} - \mathcal{P''}$. 
The correspondence for the different configurations is discussed in the following sections.

The convenience of path-dependent formalism can show itself in the semiclassical analysis where the
propagator
can
be defined via the classical action $S_c$. The associated action of a charged particle
interacting with an electromagnetic field is given by
\be
\label{em_action}
S = - mc^2 \int_{\mathcal{P}}  d \tau - \dfrac{e}{c} \int_{\mathcal{P}}
\mathcal{A}_\mu dy^\mu \, ,
\ee
with the particle's infinitesimal proper time $c d \tau = \sqrt{d y^\mu d y_\mu}
$. Here, $\mathcal{P}$ refers to any trajectory and the classical action can be
found at the classical trajectory (world line) $\mathcal{P}_c $, which minimizes
(extremizes) the action.

Eq.~(\ref{vanishing_path_integral}) implies that if one chooses the classical path for the path
dependent vector potential, $\int_{\mathcal{P}_c} \mathcal{A}_\mu (\mathcal{P}_c) dy^\mu  = 0$ is
obtained, where the classical path $\mathcal{P}_c$ satisfies the Lorentz force law
\be
\label{lorentz_force_law}
\dfrac{\del^2 y^\mu}{\del s^2} = - \dfrac{e}{c} F^{\mu \nu} \dfrac{\del
y_\nu}{\del s} \, .
\ee
The path dependent vector potential (\ref{path_dependent_potential}) for the classical
path, then, yields
\be
\label{vector_potentail_in_pdf}
\mathcal{A}_\mu (\mathcal{P}_c, x) = \dfrac{c}{e} \int_{0}^{1}  \dfrac{\del^2
y_\nu}{\del s^2}
\dfrac{\del y^\nu}{\del x^\mu}  d  s \, ,
\ee
where the dependence on the electromagnetic fields contains only in the definition of the classical path via Eq.
(\ref{lorentz_force_law}). This simplifies the semiclassical propagator to
\be
K_{c} \sim \exp\left( - \dfrac{i \, mc^2}{\hbar} \int_{\mathcal{P}_c}  d \tau 
\right)\, .
\ee

\section{The quantization of the Nonintegrable Phase}
\label{quantization_of_phase}

\subsection{General consideration}

\begin{figure}
\centering
\includegraphics[scale=0.6]{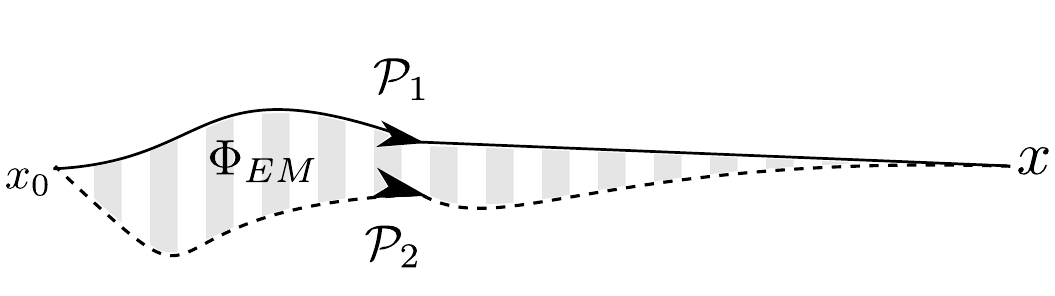}
\caption{Existence of a constant and uniform electromagnetic flux $\Phi_{EM}$ through the loop
$\mathcal{P}_1 - \mathcal{P}_2$
leads to the quantization of the nonintegrable phase.}
\label{path_phase_quantization}
\end{figure}

As we stated that the nonintegrable phase factor gives a complete description of
the true electromagnetism, let us discuss how it leads to the topological
electromagnetic effects in quantum mechanics. Consider two arbitrary paths $\mathcal{P}_1$ and
$\mathcal{P}_2$ as shown in Fig.~\ref{path_phase_quantization}, which have
the same starting
and terminating points.
Using Eq. (\ref{path_transformation}) and Eq.~(\ref{wilson_loop}), the path dependent vector potential
defined on $\mathcal{P}_2$ can be written
\be
\label{general_potential_trans}
\mathcal{A}_\mu (\mathcal{P}_2, x) = \mathcal{A}_\mu (\mathcal{P}_1, x) +
\del_\mu \Phi_{EM}(x).
\ee
If the electromagnetic flux $\Phi_{EM}$ which is defined
on the surface bounded by the loop $\del \Sigma = \mathcal{P}_1 -
\mathcal{P}_2$ is constant in both space and time, then the path dependent vector 
potentials $\mathcal{A}_\mu (\mathcal{P}_1, x)$ and $\mathcal{A}_\mu
(\mathcal{P}_2, x)$ coincide. Since the vector potential already overdescribes the true
electromagnetism \cite{Wu_Yang_1975}, i.e., different vector potentials can describe the same
physics, the wave function for a given vector potential has to be unique. As a consequence the wave
function defined on the path 
$\mathcal{P}_2$
\be
\label{constant_phase_trans}
\Psi[\mathcal{P}_2] = \exp\left(\dfrac{i e}{\hbar c} \Phi_{EM} \right)
\Psi[\mathcal{P}_1],
\ee
should match with $\Psi[\mathcal{P}_1]$ \footnote{Here it should be stressed that the constant phase
appears in Eq.(\ref{constant_phase_trans}) is a nontrivial phase based on the requirement of the
local gauge invariance. On the contrary, the trivial phase of the global gauge invariance $\chi$ can
be set equal to zero without loss of generality.}. 
Then it follows that the nonintegrable phase has to be quantized as
\be
\label{phase_quantization}
\dfrac{e \Phi_{EM}}{\hbar c} = 2 \pi \, n, \quad n= \pm 1,\pm 2, \ldots \, ,
\ee
which can be interpreted as the least condition of the flux quantization \footnote{In general,
further quantization conditions can be obtained via imposing other conditions like periodicity of
the wave function.}.

The validity of such a requirement can be further confirmed in the following way.
If there exists a constant and uniform flux $\Phi_{EM}$, then it is also possible to find
another path $\mathcal{P}_3$ whose winding number $N$ is greater than 1, i.e., a path which can wrap
the flux more than one time such that each turn can be defined on different hypersurfaces. In this
case the wave function defined for the path $\mathcal{P}_3$
\be
\Psi[\mathcal{P}_3] = \exp\left(\dfrac{i e}{\hbar c} N \Phi_{EM} \right)
\Psi[\mathcal{P}_1]
\ee 
would depend on the winding number $N$, which is inconsistent with physical realization unless
there exists an experiment which can differentiate the winding number $N$.

\subsection{Examples}
 
Let us illustrate the quantization condition for the nonintegrabel phase  in the following examples. In a certain field configuration, we choose two  paths which provide the same vector potential and show that the electromagnetic flux confined by these paths is constant and uniform and, therefore, should be quantized.
\begin{figure}
\centering
\includegraphics[scale=0.6]{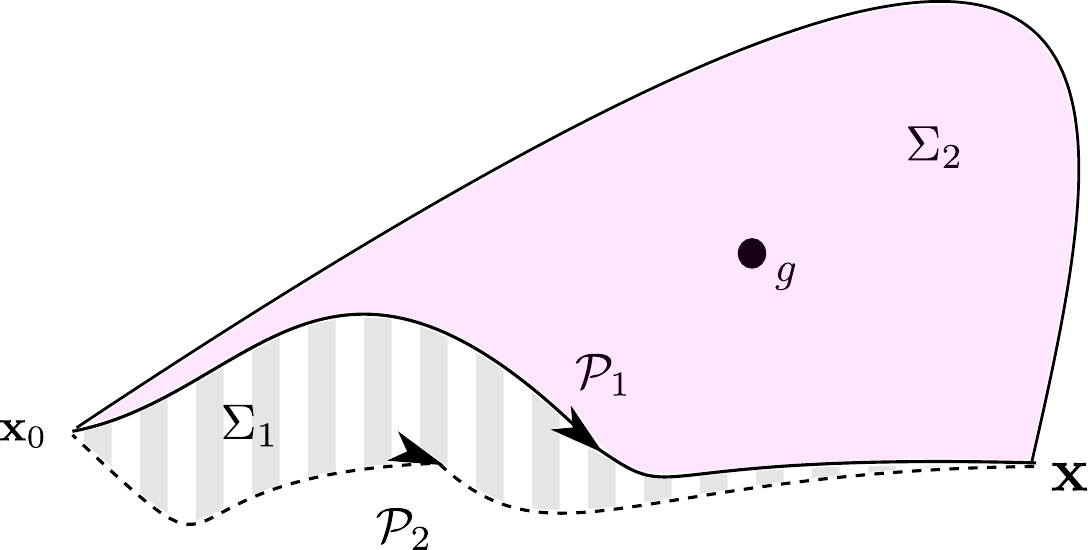}
\caption{Geometric representation of the charge quantization; when the paths
$\mathcal{P}_1$ and $\mathcal{P}_2$ overlap each other, the magnetic flux 
through the surface $\Sigma_2$ is $4 \pi g$.}
\label{path_monopole_quantization}
\end{figure}

As a first application of the electromagnetic flux quantization law for the 
constant and uniform flux Eq.~(\ref{phase_quantization}), we derive the
Dirac's charge quantization condition from the latter.
Although quantum mechanics does not require the existence of magnetic monopoles, it does not also
prohibit its presence even in the current formulation of the electromagnetism. The fundamental
relation, as it stands, $\vec{B} = \veg{\nabla} \times \vec{A}$ with a non singular free vector
potential $\vec{A}$ allows to modify the associated Maxwell equation to $\veg{\nabla} \cdot \vec{B}
= 4 \pi \, \rho_m $ with the magnetic monopole charge density $\rho_m$. Further, it was shown by
Dirac
that if there exists a magnetic monopole, it would explain why the electric charges in the nature
are
quantized \cite{Dirac_1931, Dirac_1948}. Dirac's original derivation was based on the singular
vector potential whose singularity corresponds to the so-called Dirac's string. Later, the same
result was obtained in \cite{Wu_Yang_1975} using a nonsingular vector potential defined on a
domain which is divided into two overlapping regions. In the path-dependent formalism such a
derivation was done in \cite{Cabibbo_Ferrari_1962} where the surface invariance of the closed path
is used.

Here, we will derive the same condition using the flux quantization condition
Eq.~(\ref{phase_quantization}). Consider two
paths $\mathcal{P}_1$ and $\mathcal{P}_2$ whose initial and final points coincide,
as shown in Fig. \ref{path_monopole_quantization}. Each path generates the associated vector
potential of the magnetic field due to a magnetic monopole with charge $g$. Then, there exists a
surface $\Sigma_2$ which encloses the magnetic monopole $g$. Further, if one of the paths, say
$\mathcal{P}_2$, is deformed in a way that the two paths overlap each other, then the surface
$\Sigma_1$ vanishes, and $\Sigma_2$
turns into a closed surface whose associated electromagnetic flux becomes $4 \pi g $ \footnote{Note
that the vanishing surface $\Sigma_1$ automatically satisfies the uniqueness of the wave function.}.
Consequently, stemming from Eq. (\ref{phase_quantization}), the Dirac charge quantization condition
is obtained:
\be
\label{dirac_quantization_condition}
\dfrac{2 e \, g}{\hbar c} = n.
\ee

\begin{figure}
\centering
\includegraphics[scale=0.6]{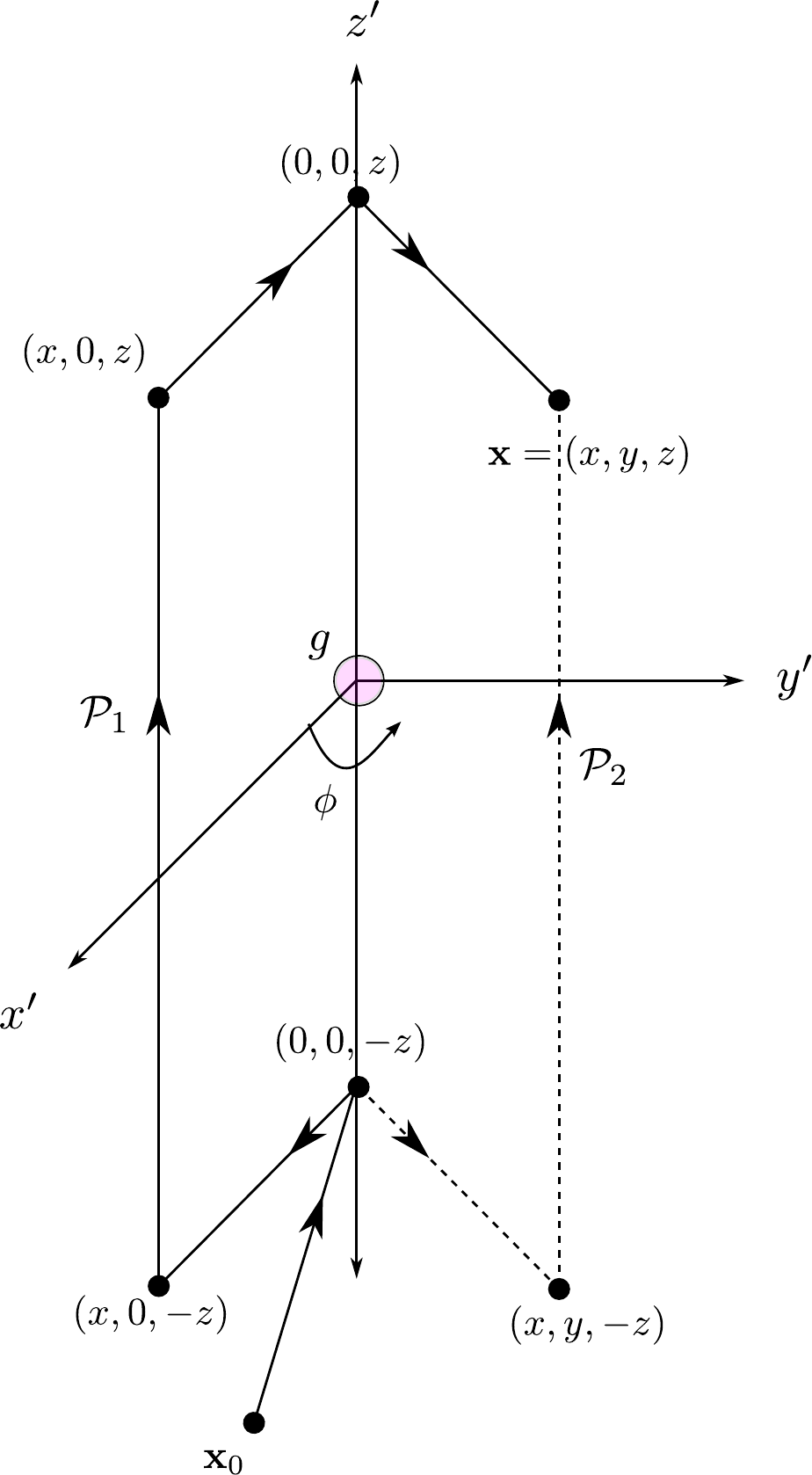}
\caption{Explicit paths which generate the associated vector potential for a
magnetic monopole.}
\label{path_monopole}
\end{figure}

Explicitly, let us take the monopole located at the origin $\vec{x} = 0$, then the magnetic field
becomes
\be
\vec{B}(r) = \dfrac{g \hat{r}}{r^2}.
\ee
If one chooses the path $\mathcal{P}_1$, see Fig. (\ref{path_monopole}), which
connects the points
\be
\vec{x}_0  \rightarrow (0,0,-z) \rightarrow (x,0,-z) \rightarrow (x,0,z)
\rightarrow (0,0,z) \rightarrow \vec{x}
\ee
with the starting point $\vec{x}_0$, say $\vec{x}_0 = (\infty,0,-\infty)$, where the vector
potential is zero, and terminating point $\vec{x} = (x,y,z)$, then the associated vector potential
becomes
\be
\vec{\mathcal{A}}(\mathcal{P}_1,\vec{x}) = \dfrac{g}{x^2+y^2} \left( y(z/r
-1),x(1-z/r),0 \right),
\ee
whose singularity corresponds to the famous Dirac's string along the half-axis
$x=y=0, z > 0$ \cite{Dirac_1931}. In spherical coordinate, it reads the convention used in
\cite{Wu_Yang_1975} 
\be
\vec{\mathcal{A}} (\mathcal{P}_1,\vec{x}) = \vec{A}^N (\vec{x}) = \dfrac{g
(1-\cos(\theta))}{r \sin(\theta)} \hat{\phi},
\ee
which is valid on the north hemisphere $0 \le \theta \le \pi - \epsilon$. Nevertheless, the path
$\mathcal{P}_2$ which connects the following points
\be
\vec{x}_0 \rightarrow (0,0,-z) \rightarrow (x,y,-z) \rightarrow \vec{x}
\ee
yields the vector potential 
\be
\vec{\mathcal{A}}(\mathcal{P}_2, \vec{x}) = \dfrac{g}{x^2+y^2} \left( y(1+z/r),-
x(1+z/r),0 \right),
\ee
whose counterpart in spherical coordinate corresponds to the vector potential
defined on the south hemisphere,
\be
\vec{\mathcal{A}}(\mathcal{P}_2, \vec{x}) = \vec{A}^S (\vec{x}) = -\dfrac{g (1 +
\cos(\theta))}{r \sin(\theta)}, \quad \epsilon \le \theta \le \pi.
\ee
Then the electromagnetic flux on the surface bounded by these two paths
\be
\label{em_flux_monopole}
\Phi_{EM} (\vec{x}) = \int_{\Sigma} \vec{B} \cdot d \vec{a} = 2 g \phi,
\ee
where the surface $\Sigma$ is the surface of a $\phi$ sphere slice.

Furthermore, as illustrated in Fig. \ref{path_monopole}, if we increase the
angle $\phi$ to $2 \pi$ which corresponds to choosing the path $\mathcal{P}_1$ as
\be
\vec{x}_0 \rightarrow (0,0,-z) \rightarrow (x,y,-z) \rightarrow (x,y,z)
\rightarrow (0,0,z) \rightarrow \vec{x},
\ee
two paths coincide and the surface of a $\phi$ sphere slice $\Sigma$ becomes the surface
of a full sphere which leads to a constant flux $\Phi_{EM} = 4 \pi g$. Then one obtains the Dirac's
charge quantization condition Eq.~(\ref{dirac_quantization_condition}). We should stress
that in this method, we have used neither periodicity, nor single-valuedness of the wave function. In
addition, we did not need further quantization conditions like quantization of the angular momentum
or the energy. 
\begin{figure}
\centering
\includegraphics[scale=0.6]{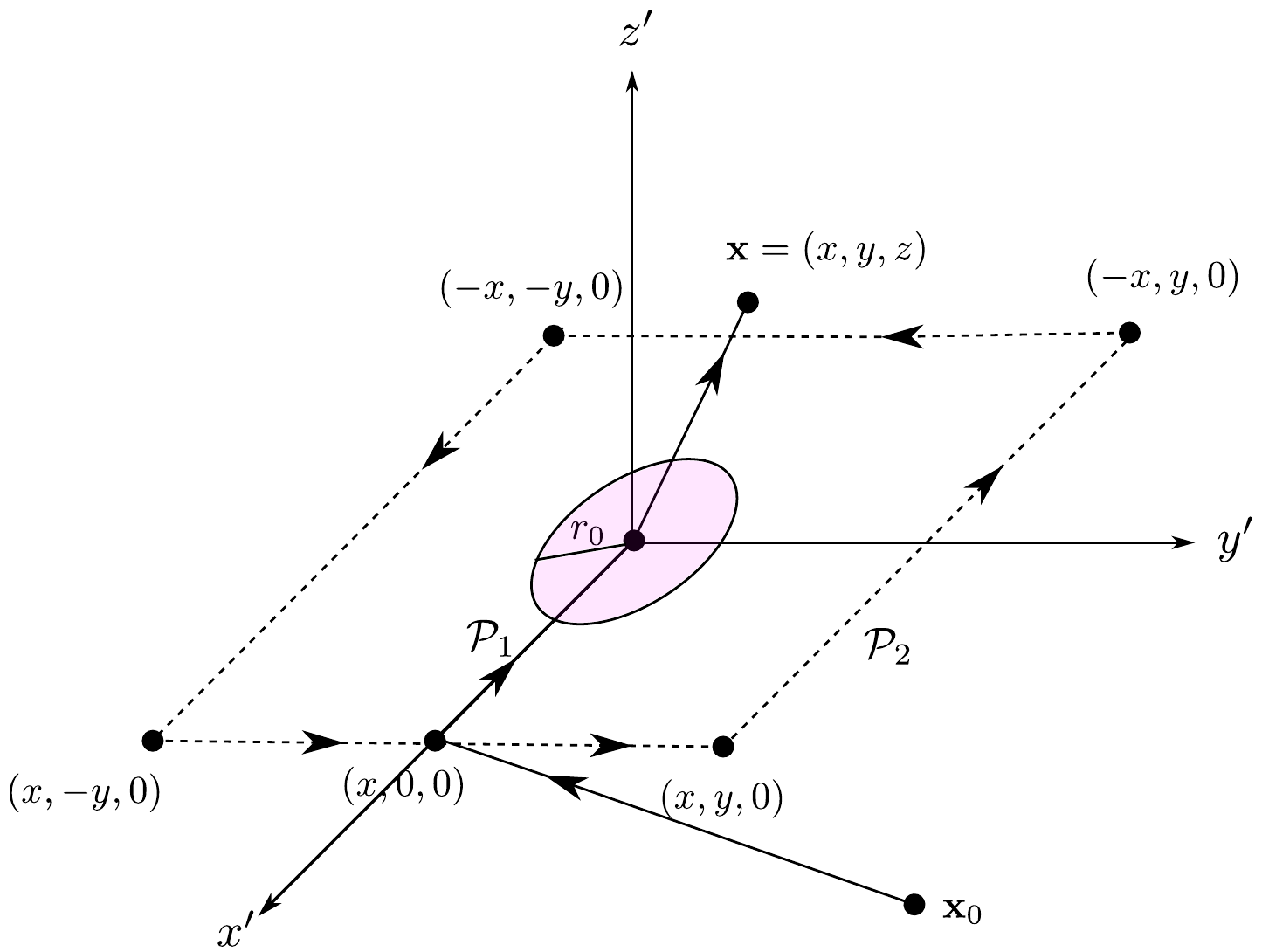}
\caption{Geometric configuration for the flux quantization. Both the path
$\mathcal{P}_1$ (solid) and the path $\mathcal{P}_2$ (dashed) give the same
vector potential. As a consequence, the flux enclosed by the loop has to be
quantized.}
\label{path_flux_quanta}
\end{figure}

As a second example, let us consider a confined static magnetic field 
\be
\vec{B}(\vec{x}) = B_0 \left(1 - \theta(r-r_0) \right) \hat{z} \, ,
\ee
with $r = \sqrt{x^2+y^2}$, $r_0=\sqrt{x_0^2+y_0^2}$, and the Heaviside step function $\theta(x)$. It
is clear that there exist two paths whose enclosed
loop give the following electromagnetic flux $ \Phi_{EM} =  B_0 \pi r_0^2 $. Then, the quantization
condition Eq.~(\ref{phase_quantization}) becomes,
\be
\label{flux_quantization_condition}
\dfrac{e B_0 \pi r_0^2}{\hbar c} =  2 \pi \, n
\ee
which corresponds to the flux quantization phenomenon in the superconducting
rings \cite{London_1950}. Specifically, one of the mentioned paths $\mathcal{P}_1$ can be chosen to 
connect the points
\be
\vec{x}_0 \rightarrow (x,0,0) \rightarrow (0,0,0) \rightarrow \vec{x},
\ee
with a starting point $x_0 = (\infty,\infty,0)$ at which the vector potential is
zero, then the path dependent vector potential yields
\be
\label{flux_quantization_potential}
\vec{\mathcal{A}}(\mathcal{P}_1,\vec{x}) = \left\{ 
  \begin{array}{l l}
    B_0 \left(-y,x,0
\right)/2, & \quad r \le r_0, \\
& \\
    \dfrac{B_0 r_0^2}{2 r^2} \left(-y,x,0
\right), & \quad r > r_0\, .
  \end{array} \right.
\ee
Similarly, as it is shown in Fig. \ref{path_flux_quanta}, the path $\mathcal{P}_2$ connecting the 
points
\bea
\nonumber \vec{x}_0 &  \rightarrow&  (x,0,0) \rightarrow (x,y,0) \rightarrow
(-x,y,0) \rightarrow (-x,-y,0) \\
&\rightarrow& (x,-y,0) \rightarrow (x,0,0) \rightarrow (0,0,0) \rightarrow
\vec{x}
\eea
will also give the same vector potential Eq.~(\ref{flux_quantization_potential}).
Although these two paths give the same vector potential, there is a non zero magnetic flux in
the surface bounded by these paths, which appears as a constant phase in the wave
function defined on the path $\mathcal{P}_2$. In order to satisfy the uniqueness
of the wave function, the phase and therefore the flux of the confined static magnetic field has to be quantized,
which gives the condition Eq.~(\ref{flux_quantization_condition}).

If the confined magnetic field, on the other hand, depends on the coordinate 
$z$ or the time (for instance its amplitude may vary in time) then the flux, in general, does not
has to be quantized due to the fact that Eq.~(\ref{general_potential_trans}) implies different
vector potentials for different paths. Note that in the experiment of the Aharanov-Bohm effect
 the existence of the confined magnetic field is realized \cite{Aharonov_Bohm_1959}. 


In the third example, we consider a constant and uniform electric field along $x$-direction, which is confined on 
a specific region of the spacetime as
\begin{align}
\vec{E}(t,x) &= E_0 \left(\theta(c t)-\theta(c t-c \Delta t)\right)\left(\theta(x)-
\theta(x-\Delta x)\right) \hat{x}.
\end{align}
Then, 
there exist two paths whose enclosed area includes the
confined electric field, which
yields the electromagnetic flux 
\be
\Phi_{EM} = c \int_{\Sigma} \vec{E}(t', x') \, d x' d t' =c \, E_0 \Delta x \Delta t.
\ee
\begin{figure}
\centering
\includegraphics[scale=0.6]{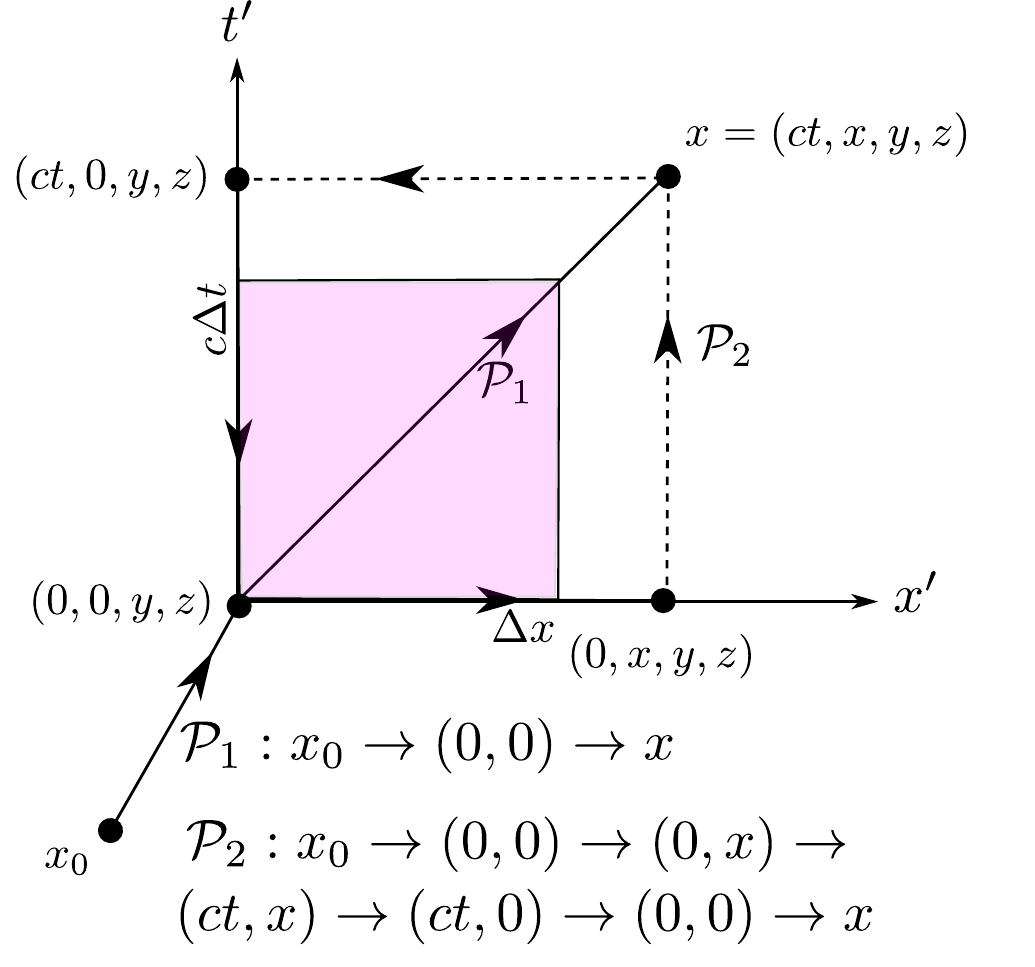}
\caption{Geometric configuration for the electric flux quantization.}
\label{path_electric_flux_quanta}
\end{figure}
Following Fig. \ref{path_electric_flux_quanta}, both the path $\mathcal{P}_1$ and the path
$\mathcal{P}_2$ give the same potential
\be
\label{electric_flux_quantization_potential}
\mathcal{A}^\mu = \left\{ 
  \begin{array}{l l}
    -E_0 \left(-x,c t,0,0
\right)/2, & \quad  \Delta x \ge x \ge 0 \wedge \Delta t \ge t \ge 0, \\
& \\
    -\dfrac{E_0 \Delta x^2 }{2} \left(\dfrac{1}{x},\dfrac{c t}{x^2},0,0
\right), & \quad x > \Delta x > 0 \wedge \Delta t > t > 0, \\
& \\
    -\dfrac{E_0 \Delta t^2 }{2} \left(\dfrac{x}{t^2},\dfrac{c}{t},0,0 
\right), & \quad \Delta x >  x > 0 \wedge t > \Delta t > 0. \\
  \end{array} \right.
\ee
Since there is a non-zero electric flux in the loop enclosed by the paths, the phase has to be
quantized such that
\be
\dfrac{e \, c \, E_0 \Delta x \Delta t}{\hbar \, c} = 2 \pi \, n
\ee
holds. Further, if the given electric field $E_0$ depends on other coordinates $y$ and/or $z$,
then the flux does not has to be quantized, hence there may exist an experiment which detects the
phase.

\section{On the electric charge quantization} \label{charge_quantization_section}

The quantization of the nonintegrable phase in the presence of the magnetic
monopole, as we have shown in the previous section, explains why the electric charges are quantized.
However, quantum mechanics cannot require magnetic monopoles to exist. Furthermore, since the
Dirac's 1931 paper which predicts magnetic monopoles to be able to quantize the electric charges,
there has never been reproducible evidence for the existence of magnetic
monopoles \cite{Particle_data_2012}. If magnetic
monopoles do not exist, then how could one explain the charge quantization? In general, not only a
magnetic monopole, but also the existence of the fundamental value for the constant and uniform confined electromagnetic flux, independent of the field configuration, would explain the charge quantization. In terms of
mathematical jargon, if there exists a non simply connected region in spacetime, then the phase can
be quantized.

Let us discuss the charge quantization in a (1+1) dimensional spacetime world.
The Maxwell equations in an arbitrary (d+1) dimensional spacetime is given \cite{Zwiebach_st}
\bea
\label{non_hom_me_arb}
 \del_\mu F^{\mu \nu} &=& \frac{2 \pi^{d/2} }{\Gamma(d/2)} \frac{J^\nu}{c} \, , \\
\label{hom_me_arb}
 \epsilon^{\alpha \beta \mu \nu} \del_\mu F_{\alpha \beta} &=& 0\, ,
\eea
with the Gamma function $\Gamma(x)$. The causal electric field of a point charge $q$ moving on an arbitrary world line
$r^\mu (\tau) = \left(r^0 (\tau),r^i (\tau)\right)$ can be found via solving the Maxwell
equation~(\ref{non_hom_me_arb}) with the retarded propagator \cite{Jackson_ced}. In a
(1+1) dimensional spacetime, then, the causal electric field can be written as
\be
\label{one_dim_electric_field}
E (t,x) = q \left( \theta\left(x - r(\tau_{+}) \right) - \theta\left(r(\tau_{-})-x \right)
\right)\, ,
\ee
where the retarded time is given by $ct - r^0(\tau_\pm)= \pm \left( x - r^1(\tau_\pm) \right)$. 

\begin{figure}
\centering
\includegraphics[scale=0.6]{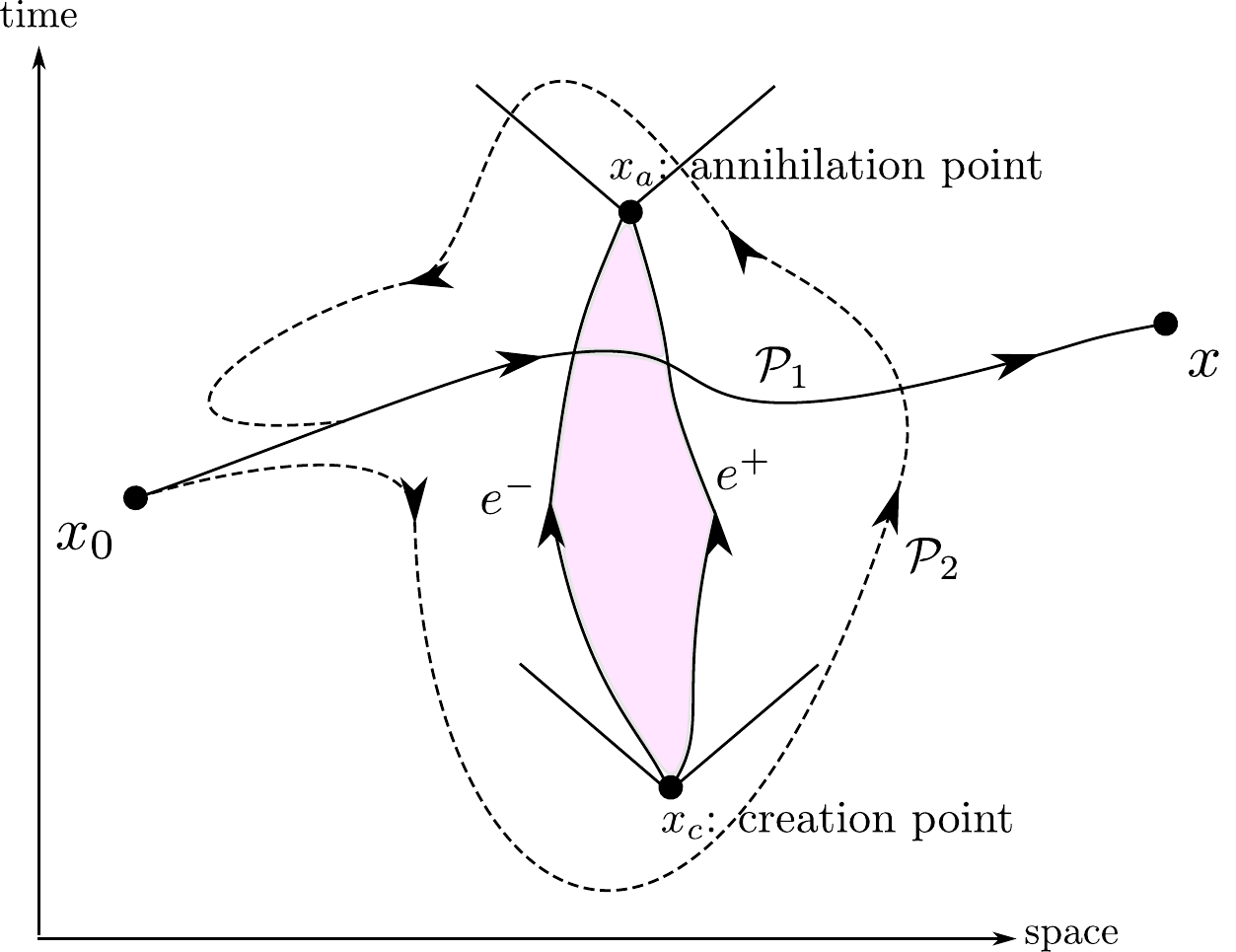}
\caption{The spacetime are bounded by the world line of each pair implies a constant uniform confined
electric field in a (1+1) dimensional spacetime, which leads to the phase quantization.}
\label{pair_production}
\end{figure}

In addition to the (1+1) dimensional causal spacetime, let us assume that there exists a quantum
vacuum
with a possibility of virtual creation and annihilation of electron-position pairs. Then, due to
an electron-positron pair creation in this universe, a confined electric field arises on the
spacetime region bounded by the world lines of the each pair as shown in Fig.~\ref{pair_production}.
Consequently, the flux defined on the spacetime area $A$ bounded by the world line of the electron
and the world line of the positron reads
\be
\Phi_{EM} = 2 \, e \, A
\ee
with the charge of the electron $e$. Then, the flux quantization condition implies
\be
\label{1d_quantization}
\alpha^{(1)}A = \pi \, n,
\ee
where $\alpha^{(1)}$ is the fine structure constant for the (1+1) dimensional spacetime. There are
many ways to read the condition of Eq.~(\ref{1d_quantization}). First of all, if the area $A$ has
a fundamental value which is determined by the universe, then the condition of Eq.~(\ref{1d_quantization})
not only could explain why the charge is quantized in the universe, but also would predict the
fundamental units of the charge. Secondly, in the other way around, the area $A$ bounded by the
world lines of the pair has to take discrete values, which leads to the existence of only discrete
set of allowed pair wordlines.

Moreover, the condition (\ref{1d_quantization}) may also be regarded as an expression which
estimates the fine structure constant for the (1+1) dimensional spacetime. Namely, the
area $A$ is bounded as $A < c^2 \tau^2 / 2$ with the lifetime of the lightest charged particle
$\tau$. Further, using the Heisenberg uncertainty relation
\be
\Delta t \,  \Delta E \ge \frac{\hbar}{2} \, ,
\ee
the flux quantization condition (\ref{1d_quantization}) estimates
\be
 \alpha^{(1)}
 \sim \lambda_C^{-2}
 \label{fine}
\ee
with
the Compton wavelength $\lambda_C = \hbar / (m_e c)$, as
long as the
Planck constant $\hbar$, the speed of light $c$, the mass of the electron $m_e$, and the fundamental
charge $e$ remain same in the (1+1) dimensional world.

It would be very remarkable to test the derived scaling law for the fine structure constant of the
one dimensional world, Eq. (\ref{fine}),
in an effectively one-dimensional solid layers like
quantum wires.

Since the world is at least (3+1) dimensional in spacetime,
the explanation of the charge quantization is missing in the absence of a magnetic monopole. 
Nonetheless, we give us a liberty to briefly speculate the possible reasons of the electric charge
quantization in a (3+1) dimensional world.
Note that the existence of 
a confined constant and uniform field anywhere in the universe would explain the charge quantization everywhere. If we
imagine that the spatial dimensions are emergent in a row from the Big Bang, instead of at the same
time, one could think that
the charge is quantized due to the constant uniform confined
electromagnetic field in the (1+1) spacetime structure of the early universe.

\section{Conclusion} \label{conclusion}

The nonintegrable phase factor which describes the complete theory of
electromagnetism can replace the gauge freedom on the vector potential with the path freedom.
In this equivalent formulation of the gauge theory, we have shown the quantization of the electromagnetic
flux generated from the constant and uniform confined field. This would imply and explain the charge quantization,
if the existence of the fundamental confined field with a uniform and constant flux is proven.
Finally, in the absence of the evidence for magnetic monopoles, we are naively questioning the possible
reasons of the charge quantization via investigating the quantization condition of the nonintegrable
phase. It was shown that the (1+1) dimensional world could allow to explain the quantization of the
charge as well as the estimation of its fundamental unit.

\section*{Acknowledgments}

We are grateful to C. H. Keitel, J. Evers,  M. Klaiber, S. Meuren, and A. W\"{o}llert for valuable
discussions.

\bibliography{fundamental_qm_bib}

\end{document}